\documentstyle[aps,multicol,epsfig]{revtex} 
  
\begin{document}  
\title{Magnetic field correlations in kinematic   
two-dimensional magnetohydrodynamic turbulence}  
\author{\sc J\"org Schumacher and Bruno Eckhardt }  
\address{Fachbereich Physik, Philipps-Universit\"at Marburg,  
	   D-35032 Marburg, Germany}  
\date{November 1999}  
\maketitle  
  
\begin{abstract}  
The scaling properties of the second order magnetic structure function 
$D_2^{(B)}(r)$ and the corresponding magnetic correlation function $C_2^{(B)}(r)$ 
are derived for two-dimensional magnetohydrodynamic turbulence in the kinematic 
regime where the ratio of kinetic energy to magnetic energy is much larger than 
one.  In this regime the magnetic flux function $\psi$ can be treated as a passive 
scalar advected in a two-dimensional turbulent flow.  Its structure function 
$D_2^{(\psi)}(r)$ and the one for the magnetic field $D_2^{(B)}(r)$ are connected 
by an exact relation.  We calculate $D_2^{(\psi)}(r)$ and thus $D_2^{(B)}(r)$ 
within geometric measure theory over a wide range of scales $r$ and magnetic 
Prandtl numbers $Pr_m$.  The magnetic field correlations follow a 
$r^{-4/3}$--scaling law and show an anticorrelation at the beginning of the 
Batchelor regime indicative of the formation of strongly filamented 
current sheets.  Differences to the full dynamic 
regime, where the ratio of kinetic to magnetic energies is smaller than  
in the kinematic case, are discussed. 
\end{abstract}  

\begin{multicols}{2}
\section{Introduction}  
Magnetohydrodynamic (MHD) turbulence appears in many problems in nuclear fusion 
research as well as in astrophysics and geophysics (see, e.g.  Refs.  1 and 2 
for reviews).  Compared to the already complicated hydrodynamic turbulence there 
are additional degrees of complexity due to the presence of a magnetic field, 
the possibility of dynamos and the action of the magnetic field back on the 
flow through the Lorentz force.  In addition, the distribution of  
energy among the modes is controlled through the interaction of 
three cascade processes, related to the three conserved quantities 
of the dynamical equations in the inviscid limit: the total energy, the cross 
helicity, and for two dimensions the mean square magnetic potential. 
There is, however, a regime where some of the more complicated processes are 
subdominant or absent and where a rather complete analysis can be performed: 
two-dimensional (2-D) magnetohydrodynamics with weak magnetic fields. 
This regime can be relevant to astrophysical applications,  
such as quasi-two-dimensional turbulent processes in thin accretion  
discs.\cite{Kui95} 
 
The situation we consider is that of a 2-D turbulent flow, maintained by some 
external force, in which a magnetic field is advected.  The linearity of the 
equation for the magnetic induction implies that without an external seed field 
no magnetic field will ever be generated.  Moreover, the absence of dynamo 
action in two dimensions \cite{Zel80} implies that without permanent driving the 
magnetic field will die out eventually.  The action of the magnetic field back 
on the flow can be neglected if the Lorentz force is sufficiently weak, i.e., if 
the magnetic field is sufficiently small.  Because of the absence of dynamos in 
2-D the amplitude of the external driving 
controls the magnitude of the magnetic field.
We thus have a regime of kinematic evolution of  
a magnetic field passively advected 
by the turbulent flow which can be realized in any 2-D situation 
for sufficiently small driving of the magnetic field. 
 
The dynamics of the magnetic field and thus its correlation functions are 
nevertheless nontrivial since a transient growth and the development of fractal 
like structures on small scales are possible, as evidenced by numerical 
simulations.  \cite{Bis90,Cat91} Though the range of applicability of this 
kinematic approach is limited, it shows some interesting features which we would 
like to describe here. 
 
On a technical level, the restriction to a 2-D situation allows to connect the 
evolution of the vector potential to that of a scalar field for which detailed 
scaling predictions exist within geometric measure 
theory. \cite{Con93,Proc93,EckSch98} 
On the assumptions of statistical stationarity, homogeneity, and isotropy for 
both the magnetic field and the scalar field, geometric measure theory provides 
a connection between the structure functions of the velocity field and the 
scalar field.  In particular, the scaling behavior can be related and the 
dependence on the Prandtl number and on another parameter that characterizes the 
strength of the driving and appears in 2-D only can be extracted.  Using an 
exact relation between the correlation function of the scalar and the magnetic 
field we can transfer these results to the magnetohydrodynamic problem and in 
particular discuss the dependence on the magnetic Prandtl number.  The latter 
can range from $Pr_m\sim 10^{-6}$ in the solar convection zone to $Pr_m\sim 
10^{7}$ in the interstellar medium. \cite{Zel83}  Within the kinematic regime we 
are able to go beyond previous applications of geometric measure theory to MHD 
turbulence. \cite{Bis93a,Gra95} 
 
The outline of the paper is as follows.  In sections II and III the theoretical 
background in 2-D magnetohydrodynamics and 2-D passive scalar advection is 
summarized.  For details on the analysis of the 2-D passive scalar within 
geometric measure theory we refer to our previous work. \cite{EckSch98}  In 
section IV both are combined and the correlation and structure functions of 
magnetic fields in the kinematic regime are discussed.  In section V we compare 
with correlation functions and structure functions found in full numerical 
simulations; this highlights features of the interaction between both fields 
which have to be included when the magnetic field becomes dynamically relevant. 
We conclude with a summary in section VI.

\section{MHD basics}  
The equations for a magnetic field ${{\bf B}}({\bf x},t)$ in a conducting  
fluid moving with velocity ${\bf u}({\bf x},t)$ are 
\end{multicols} 
\begin{eqnarray}  
\rho\left(\partial_t {\bf u} + ({\bf u}\cdot\nabla){\bf u}\right) &=& -\nabla p +  
\frac{1}{\mu_0}({\bf\nabla\times}{\bf B})\times{\bf B}  
+\rho\nu\Delta{\bf u} + \rho{\bf f}_u \\  
\partial_t {\bf B} + ({\bf u}\cdot\nabla){\bf B} &=& ({\bf B}\cdot\nabla){\bf u} +   
\eta \Delta {\bf B} + {\bf f}_B \\  
\nabla\cdot {\bf u} &=& \nabla \cdot {\bf B} = 0  
\end{eqnarray} 
\begin{multicols}{2} 
where $\nu$ is the kinematic viscosity, $\eta$ is magnetic diffusivity, 
$p$ is the thermal pressure, and 
${\bf f}_u$ and ${\bf f}_B$ are the external driving for each field.  The quantity $\rho$ 
is the constant mass density and $\mu_0$ denotes the permeability in a vacuum. 
We estimate typical amplitudes for the velocity field and the magnetic field by 
the root mean square values $u_{r.m.s.}=\langle u_x^2\rangle^{1/2}$ and 
$B_{r.m.s.}=\langle B_x^2\rangle^{1/2}$ where $\langle\cdot\rangle$ denotes  
a statistical  
average over an isotropic and homogeneous ensemble.  
If then $L$ is an external length 
scale, the fluid Reynolds number $R=u_{r.m.s.}  L/\nu$ and the magnetic Reynolds 
number $R_m=u_{r.m.s.}  L/\eta$ can be formed. Their ratio defines 
the magnetic Prandtl number $Pr_m=\nu/\eta=R_m/R$. The relative size 
of magnetic field and velocity field is measured by the  
Alfv\'en-Mach number $M$, the ratio 
of the typical fluid velocity $u_{r.m.s.}$ to the Alfv\'en velocity 
$v_A=B_{r.m.s.}/(\mu_0\rho)^{1/2}$,  
\begin{eqnarray}  
M&=&\frac{u_{r.m.s.}}{v_A},\nonumber\\ 
 &=& \left(\frac{\mu_0\rho u_{r.m.s.}^2}{B_{r.m.s.}^2}\right)^{1/2},\nonumber\\ 
 &=&\left(\frac{E_K}{E_B}\right)^{1/2}\,.   
\end{eqnarray} 
By the last line this is also the square root of the ratio between  
fluid kinetic energy and magnetic field energy.  
 
With these definitions, the dimensionless form of  
the first two equations reads 
\end{multicols} 
\begin{eqnarray}  
\partial_t {\bf u} + ({\bf u}\cdot\nabla){\bf u} &=&  
-\nabla p  
+M^{-2}({\bf\nabla\times}{\bf B})\times{\bf B}  
+R^{-1}\Delta{\bf u} +  
{\bf f}_u \\  
\partial_t {\bf B} + ({\bf u}\cdot\nabla){\bf B} &=& ({\bf B}\cdot\nabla){\bf u} +   
R_m^{-1} \Delta {\bf B} + {\bf f}_B \,.  
\end{eqnarray}  
\begin{multicols}{2} 
In two dimensions, both the velocity field and the magnetic field can be  
represented by scalar fields by introducing the vorticity  
\begin{equation}  
\omega {\bf e}_z={\boldmath\nabla\times}{\bf u}\,,  
\end{equation}  
and the magnetic flux function 
\begin{equation}  
{\bf B}={\boldmath\nabla}\psi{\boldmath\times}{\bf e}_z  
\,.  
\end{equation}   
The MHD equations then take on the form  
\begin{eqnarray}  
\label{nseq_nondim}  
\partial_t\omega  
 +({\bf u}\cdot{\bf\nabla})\omega  
 &=&  
 M^{-2}({\bf B}\cdot{\bf\nabla})j  
 +R^{-1}{\bf\nabla}^2\omega  
 +f_{\omega}\,,\\  
\label{indeq_nondim}  
\partial_t\psi  
 +({\bf u}\cdot{\bf\nabla})\psi  
 &=&   
 R_m^{-1}{\bf\nabla}^2\psi  
 +f_{\psi} \,,  
\end{eqnarray}  
with $j=-\nabla^2\psi$  the 
$z$-component of the current density 
and $f_{\psi}$ and $f_\omega$ the scalar driving fields .  
  
The situation we want to consider is one of weak magnetic fields, i.e.  very 
large $M$, so that the term with the Lorentz force density can be ignored.   
This can be estimated to be reasonable\cite{Bis90} if the Mach number exceeds the 
magnetic Reynolds number, i.e. if $M>R_m$. 
Then the equations for the fluid do not depend on 
the magnetic field any more and the statistically stationary state that develops 
depends only on the strength of the force $f_{\omega}$ and the length scale 
$l_f$ on which it acts.  In 2-D turbulence it is the enstrophy that cascades 
down to the viscous scales.  So if we assume homogeneity and isotropy the 
appropriate smallest scale is the enstrophy dissipation length 
$\eta_{\omega}=\nu^{1/2}\epsilon_{\omega}^{-1/6}$, defined in terms of the 
enstrophy dissipation rate 
$\epsilon_{\omega}=\nu\langle|{\boldmath\nabla}\omega|^2\rangle$. 
 
The quantities we want to study are the 
second order structure function $D_2^{(B)}(r)$, 
\begin{equation}   
\label{definition_D2}  
D_2^{(B)}(r)=\langle|{\bf B}({\bf x}+{\bf r})  
-{\bf B}({\bf x})|^2\rangle \,,  
\end{equation}  
and the correlation function $C_2^{(B)}(r)$,  
\begin{equation}  
\label{definition_C2}  
C_2^{(B)}(r) = \langle {\bf B}({\bf x}+{\bf r})\cdot  
				  {\bf B}({\bf x})\rangle \,,  
\end{equation}  
of the magnetic field.  $D_2^{(B)}(r)$ and  
$C_2^{(B)}(r)$ are related by 
\begin{equation}  
\label{definition_DC}  
D_2^{(B)}(r) = 2 \langle B^2 \rangle -2 
C_2^{(B)}(r)\,.   
\end{equation}  
Continuity of the fields on small scales and decay 
of correlations on large scales implies the following 
limiting behavior for 
small and large distances $r$: 
\begin{equation}  
\label{db_asympt} 
D_2^{(B)}(r)=\left\{\matrix{\epsilon_B r^2/(2\eta)\rightarrow 0 
		 &:\;\;r\rightarrow 0\cr  
		 2\langle B^2 \rangle 
		 &:\;\;r\rightarrow L\;,}\right.  
\end{equation}  
whereas it is the other way round for the correlation function,  
\begin{equation}  
C_2^{(B)}(r)=\left\{\matrix{2\langle B^2\rangle 
		 &:\;\;r\rightarrow 0\cr  
		 0 
		 &:\;\; r\rightarrow L\;.}\right.  
\end{equation}  
Similar quantities can be defined for the velocity field and the magnetic 
flux function $\psi$. The quantity  
$\epsilon_B=\eta\langle|{\bf\nabla}{\bf B}|^2\rangle$ is the 
magnetic energy dissipation rate.  Since $\psi$ obeys in the kinematic limit 
the equations of a passive scalar, we can use our previous results 
for the 
scaling of a passive scalar in a turbulent fluid to obtain the scaling regimes 
for $\psi$.  They can be transfered to the magnetic field case 
by use of the relations\cite{GroMer92} (see Appendix A for the derivation)  
\begin{eqnarray}   
\label{laplace}  
D_2^{(B)}(r)=2\frac{\epsilon_{\psi}}{\eta}-\Delta D_2^{(\psi)}(r),  
\end{eqnarray}   
and consequently with (\ref{definition_DC}) 
\begin{eqnarray}   
\label{laplace1}  
C_2^{(B)}(r)=\frac{1}{2}\Delta D_2^{(\psi)}(r).  
\end{eqnarray}  
The quantity $\epsilon_{\psi}=\eta\langle|{\bf\nabla}\psi|^2\rangle
=\eta\langle B^2\rangle$ is the mean flux 
dissipation rate.

\section{2-D passive scalar advection} 
\label{frame_geo} 
In the limit of large $M$ the magnetic 
flux $\psi$ does not act on the flow field ${\bf u}$,  
so that (\ref{indeq_nondim}) describes the  
passive advection of a scalar field in the velocity field that 
follows from the hydrodynamic driving. A connection between the  
scaling behavior of the fluid, as contained in the  
velocity structure function, 
\begin{equation} 
D_2(r)=\langle|{\bf 
u}({\bf x}+{\bf r})-{\bf u}({\bf x})|^2 \rangle\,, 
\end{equation} 
and its longitudinal part, 
\begin{equation} 
D_{\parallel}(r)= 
\frac{1}{r^2}\int_0^r \rho \,D_2(\rho) \,\mbox{d}\,\rho\,, 
\end{equation} 
and the scaling behavior of the second order scalar field structure 
function, 
\begin{eqnarray}  
D_2^{(\psi)}(r)&=&\langle|{\psi}({\bf x}+{\bf r})  
-{\psi}({\bf x})|^2\rangle\,, \nonumber\\  
&\propto& r^{\zeta_2^{(\psi)}}\,,  
\end{eqnarray}  
can be found within geometric measure theory.\cite{Con93,Proc93} 
One estimates the 
fractal dimension $\delta_g^{(1)}$ of level sets $\psi_0=\psi({\bf x})$ of the 
scalar field graph $G=({\bf x},\psi({\bf x}))$, and obtains in the absence of 
intermittency corrections, i.e. ${\zeta_2^{(\psi)}}=2 {\zeta_1^{(\psi)}}$,  
the upper bound 
\begin{equation}  
\label{ineq1} 
2 \delta_g^{(1)} \le 4 - {\zeta_2^{(\psi)}}\,.  
\end{equation}  
Building on Ref. 7 and extending the results of Ref. 8 we were able 
to obtain scale resolved and Prandtl number dependent
upper bounds for $\delta_g^{(1)}$, which turned out to
be rather sharp.\cite{EckSch98} 
If we assume that the bounds on $\delta_g^{(1)}$
are indeed equalities, (\ref{ineq1}) gives an 
upper bound on $\zeta_2^{(\psi)}$, viz.
\begin{eqnarray}   
\label{fracdim}   
{\zeta_2^{(\psi)}} = 2-   
2\frac{\mbox{d}}{\mbox{d}\,\ln r}   
\ln\sqrt{1+\alpha\,Pr_m\,r^2+\sqrt{3} Pr_m r \sqrt{D_{\parallel}}}\,.  
\end{eqnarray}   
In the following we will evaluate the right hand side and in particular
its dependence on the scaling of the velocity field and take this
as an estimate for the scaling exponent $\zeta_2^{(\psi)}$. As mentioned,
this relies
on the assumption that several inequalites are in fact equalities
and that intermittency effects are absent. A careful analysis of the
derivation shows that the presence of intermittency in the scalar
field modifies several approximations and that the cumulative effect
is difficult to control. It is fairly straightforward, however,
to allow for intermittency effects in the velocity structure function.

The dimensionless parameter 
$\alpha=\epsilon_{\psi}/(\epsilon_{\omega}^{1/3}\psi_{r.m.s.}^2)$ measures 
the ratio of the scalar driving rate $\epsilon_{\psi}/\psi_{r.m.s.}^2$ to the 
flow advection rate $\epsilon_{\omega}^{1/3}$ and is specific to two-dimensional 
passive scalar advection.  For large values of $\alpha$ the scalar cannot be 
advected sufficiently rapidly to the small dissipative scales so that a space 
filling distribution with $\zeta_2^{(\psi)}=0$ develops.  The distance $r$ is 
measured in units of $\eta_{\omega}$ and the longitudinal structure function 
$D_{\parallel}(r)$ of the velocity field is measured in units of 
$\epsilon_{\omega}^{2/3} \eta_{\omega}^2$.  The explicit dependence on the 
longitudinal velocity structure function $D_\parallel$ allows to substitute 
numerical or experimental findings for the velocity structure function, which 
very often differ from theoretical expectations, and to obtain predictions for 
the corresponding scalar structure functions.  
 
As in the previous work \cite{EckSch98} we use in particular structure functions 
obtained from Fourier transforms of model spectra in $k$-space and evaluate 
(\ref{fracdim}) numerically.  Recent experiments on forced two-dimensional 
turbulence\cite{Par97,Rut98} and a number of direct numerical 
simulations\cite{FriSul84,SmiYak93,Ben86} support the existence of a 
Kolmogorov-like scaling for the energy spectrum, $E(k)\sim\,k^{-5/3}$ in the 
inverse energy inertial subrange $k<k_f$ and a scaling $E(k)\sim\,k^{-\beta}$ 
with $\beta\ge3$ for the direct enstrophy inertial subrange $k>k_f$.  
Recent experiments on 2-D turbulence indicate that the velocity increments 
in both cascade regimes show a non-intermittent Gaussian  
behavior.\cite{Par97,Tab99}  
We do not discuss here the role of coherent, large scale vortices which are 
formed by the inverse energy cascade in two-dimensional hydrodynamic turbulence 
and which can effect the spectra in the enstrophy subrange.  This problem is 
still a matter of debate.  \cite{Bab87,Bas94} We therefore take for our analysis 
the following model spectrum for the amplitudes $\langle|{\bf u}_{\bf 
k}|^2\rangle$ of the velocity field in a Fourier representation in a periodic 
box of size $L=2\pi$: 
\end{multicols} 
\begin{eqnarray} 
\label{spectrum} 
\langle|{\bf u}_{\bf k}|^2\rangle\!\sim\! 
\left\{  
\begin{array}{r@{\quad:\quad}l}  
k_f^{-7/3} k_1^{-11/3} k^3   & \frac{2\pi}{L}\le k\le k_1, \\  
k_f^{-7/3} k^{-2/3}          & k_1<k\le k_f,\\  
k^{-3}        & k_f<k\le k_{\omega}=\frac{1}{\eta_{\omega}},\\  
k^{-3}\exp\left[-\left(\frac{k-k_{\omega}}{k_{\omega}}\right)^{2}\right]  
		    & k_{\omega}<k\;.  
\end{array} 
\right. 
\label{modspec} 
\end{eqnarray} 
\begin{multicols}{2}
Note the differences in scaling between the amplitudes 
$\langle|{\bf u}_{\bf k}|^2\rangle$ and the 
energy spectrum $E(k)$ due to phase space factor, i.e. $E(k)\!\sim\!k^{-\beta-1}$ 
corresponds to $\langle|{\bf u}_{\bf k}|^2\rangle\!\sim\!k^{-\beta}$.  
For a dependence of the 
results on $\beta$ the reader is refered to our previous work,  
where we showed that 
a variation of $\beta$ in a range between 2 and 4 left the scaling of $D_2(r)$ 
in the enstrophy intertial subrange nearly unchanged. 
We therefore take the exponent $\beta=3$ for 
$\langle|{\bf u}_{\bf k}|^2\rangle$ [see Eq. (\ref{modspec})]
in all calculations here. 
 
Assuming stationarity, homogeneity, and isotropy 
the relation between velocity spectrum scaling and the velocity structure  
function $D_2(r)$  is given  by the volume average  
\begin{eqnarray} 
D_2(r)&=&\frac{1}{V}\int_{V}\,|\,{\bf u}({\bf x}+{\bf r})-{\bf u}({\bf x})\,|  
	 ^{2}\,\mbox{d}\!V,\nonumber \\  
	&=&\frac{1}{V}\int_{V}\,|\,\sum_{{\bf k}}{\bf u}_{\bf k}\,  
	 \exp(i{\bf k}\cdot{\bf x})  
	[\exp(i{\bf k}\cdot{\bf r})-1]\,|^{2}\,\mbox{d}\!V,\nonumber \\  
	&=&2\,\sum_{{\bf k}}\,\langle|{\bf u}_{\bf k}|^{2}\rangle\,  
	 (1-\cos{({\bf k}\cdot{\bf r})})\, .  
\end{eqnarray}    
By averaging over all directions (due to isotropy) in ${\bf k}$-space   
the cosine gives rise to the Bessel function $\mbox{J}_0(kr)$,  
\begin{equation}  
\label{mod_struc}  
D_2(r)=2\,\sum_k\,\langle|{\bf u}_{\bf k}|^{2}\rangle\,  
(1-\mbox{J}_{0}(kr)) \,.  
\end{equation}  
The model spectrum (\ref{modspec}) is then substituted and the summation in 
(\ref{mod_struc}) is evaluated numerically using a finite  
set of wave numbers, equidistant  on a logarithmic scale. 
We checked the independence of our results for  
several wavenumber resolutions .  
 
The algebraic scaling of $D_2(r)$  
with respect to $r$ is shown in Fig. 
\ref{p1} (thick dash-dotted line).  For the dissipation scales  
$r<\eta_{\omega}$ it follows a $r^2$--dependence in 
correspondence with the Taylor expansion.  This quadratic scaling with respect 
to $r$ continues in the enstrophy inertial subrange for scales 
$\eta_{\omega}<r<l_f$.   
For distances $l_f<r$ the energy inertial subrange with 
a $r^{2/3}$--scaling sets in.  Finally, a saturation to a constant value due 
to finite size effects is observed.

\section{2-D kinematic MHD turbulence}  
\label{Res}  
 
Using the energy spectra (\ref{spectrum}) we can now evaluate the passive scalar 
structure and correlation function exponents and, via (\ref{laplace}), the 
corresponding behavior for the magnetic fields.  The results for a large range 
of length scales and magnetic  
Prandtl numbers are shown in Figs.  \ref{p1} and \ref{p2}. 
The structure function is plotted in units of 
$\epsilon_{\psi}\epsilon_{\omega}^{-1/3}$, and $\log_{10} D_2^{(\psi)}$ is obtained by 
numerical integration of $\zeta_2^{(\psi)}(r)$ over $\log_{10} r$.  We find in both 
figures the following scaling regimes.  For small scales $r<\eta_{\omega}$ the 
fields are smooth and one finds $D_2^{(\psi)}(r)=(\epsilon_{\psi}/2\eta) r^2$ 
and thus $\zeta_2^{(\psi)}=2$.  The quadratic scaling continues into the 
enstrophy inertial subrange $r\gtrsim\eta_{\omega}$.  For sufficiently large 
$Pr_m$ the third term under the square root in Eq.  (\ref{fracdim}) gives the 
dominant contributions and the scaling exponent changes to $\zeta_2^{(\psi)}=0$; 
this is the viscous--convective Batchelor regime\cite{Bat59}.  The flux $\psi$ 
is then advected chaotically in a velocity field that is still smooth on these 
scales, i.e.  $D_{\parallel}\sim r^2$ and the flux contours are stretched and 
twisted by the fluid motion and form filamented current sheets.  The 
$r^{2/3}$--scaling regime for larger $r$ and for both the velocity field and the 
magnetic flux function is connected with the inverse cascade of the underlying 
fluid turbulence; this is the inertial--convective Obukhov--Corrsin 
regime\cite{Obu49}.  The final saturation to a constant value for very large 
separations is due to the finite system size $L$. 
 
The scaling of the structure function $D_2^{(B)}(r)$ as well as of the 
correlation function $C_2^{(B)}(r)$ follow from Eqns.  (\ref{laplace}) and 
(\ref{laplace1}), respectively.  The results for the structure function of the 
magnetic field $D_2^{(B)}(r)$ in units of $\epsilon_{\psi}\nu^{-1}$ are shown 
in Fig.~\ref{p3} (thick solid lines).  Clearly, for scales $r<\eta_{\omega}$ the 
structure function shows again the Taylor expansion behavior 
$D_2^{(B)}(r)=(\epsilon_B/2\eta)r^2$ as stated in Eq.  (\ref{db_asympt}).  The 
larger the scales the more dominant the constant first term in Eq. 
(\ref{laplace}), which is proportional to the magnetic Prandtl number, leading 
to a saturation at a constant value $D_2^{(B)}(\infty)=2\epsilon_{\psi}/\eta$ 
for large distances, i.e.  where $D_2^{(B)}\sim r^0$.  The crossover between 
both scaling regimes is relatively sharp and shifted towards smaller $r$ for 
growing values of $Pr_m$.  We could not observe an intermediate algebraic 
scaling between both ranges. 
 
The spatial correlations in the magnetic field are subdominant in $D_2^{(B)}(r)$ 
for scales $r\gtrsim\eta_{\omega}$, but show up in the correlation  
function $C_2^{(B)}(r)$:  an algebraic scaling $D_2^{(\psi)}(r)\sim r^{\alpha}$ 
corresponds to $C_2^{(B)}(r)\sim r^{\alpha-2}$.  The results for the correlation 
function, derived using (\ref{laplace1}), are shown in Fig.~\ref{p4} for 
different values of $Pr_m$ and $\alpha$.  In the logarithmic plot the algebraic 
decay of the correlation for all discussed parameter sets is clearly visible in 
particular, a $r^{-4/3}$--decay is found in regions where $D_2^{(\psi)}(r)\sim 
r^{2/3}$ (see also Fig.~\ref{p2}).  The correlations decay very fast when the 
magnetic flux is advected passively with the turbulent velocity field. 
 
The results show an anticorrelation of the magnetic fields,  
i.e. $C_2^{(B)}< 0$. It is  
observed near the crossover from $D_2^{(\psi)}\sim r^2$ to  
$D_2^{(\psi)}\sim r^0$, i.e. near the transition from the  
viscous regime to the Batchelor plateau.  
This follows also analytically within the Batchelor  
parametrization \cite{Bat51} of $D_2^{(\psi)}(r)$ by setting   
\begin{eqnarray}  
D_2^{(\psi)}(r)=\frac{r^2}{1+a_2 r^2}\;,  
\end{eqnarray}  
which approximates such a crossover very well. Equation  
(\ref{laplace1}) then gives 
\begin{eqnarray} 
\label{lap-bat}  
\frac{1}{r}\partial_r(r \partial_r D_2^{(\psi)}(r))  
=4 \frac{1-a_2 r^2}{(1+a_2 r^2)^3}\;.  
\end{eqnarray}  
With $a_2 >0$ and $r\ge 0$ we find negative values of (\ref{lap-bat}) for 
$r>(a_2)^{-1/2}$.  It can be shown that a crossover of 
$D_2^{(\psi)}(r)$ from $r^{\alpha}$ to $r^{\beta}$ with $0\le\beta\le\alpha\le 
2$ gives negative values  
in (\ref{lap-bat}) and thus anticorrelation of the 
magnetic field at all when 
in addition the constraint $\beta<\alpha/5$ is satisfied. 
This is consistent with our results for extremely small magnetic Prandtl numbers 
where no anticorrelation is observed.  Here, the quadratic scaling with $r$ 
changes directly to an $r^{2/3}$--scaling (see Figs.~\ref{p2} and \ref{p4}), and 
the intermediate Batchelor plateau is not present. 
 
The anticorrelation in the normalized correlation function is magnified in Fig. 
\ref{p5} for values of $Pr_m$ reaching from $10^{-5}$ to $10^3$.  The negative 
contributions vanish for sufficiently low $Pr_m$ where the Batchelor plateau is 
supressed (see the cases $Pr_m=10^{-4},\,10^{-5}$).  On the other hand we 
observe a nearly constant value $\min\left(C_2^{(B)}\right)\simeq 
-0.0375\epsilon_{\psi}/\eta$ for the larger values of the magnetic Prandtl 
number. 
 
The origin of this anticorrelation presumably lies in the advection 
of $\psi$ in turbulent flow structures  
and the formation of strongly filamented flux sheets (and thus  
current sheets) in which the magnetic energy is stored. The 
minimal sheet width $\delta_{CS}$  
can be determined by the balance between the advection 
and the diffusion term of (\ref{indeq_nondim}), giving 
\begin{eqnarray}  
\frac{U_l}{l}\psi\simeq\eta\frac{\psi}{\delta_{CS}^2}\,.  
\end{eqnarray}  
By taking $U_l/l=\epsilon_{\omega}^{1/3}$ as the typical strain rate we end 
up with 
\begin{eqnarray}  
\label{delta}  
\delta_{CS}\simeq \eta_{\omega}Pr_m^{-1/2}\,.  
\end{eqnarray}  
This $Pr_m$ scaling is also found in the position of the 
minimum, so that the anticorrelation is due to layers  
with opposite orientations of magnetic fields. The saturation 
of the anticorrelation in Fig. \ref{p5} suggests that 
the ratio of regions with perfect parallel alingment and  
deviations, say due to folds or modulations in thickness,  
stays the same, independent of  
magnetic Prandtl number. 
 
The relation (\ref{delta}) can also be derived by means of (\ref{fracdim}). 
Assuming the subdominance of the $\alpha Pr_m r^2$ term, we can ask when the 
last term under the square root in (\ref{fracdim}) will exceed unity.  By 
requiring a smooth flow in the viscous subrange, i.e.  setting 
$D_{\parallel}\sim r^2$, we get the same $Pr_m$ dependence as in (\ref{delta}).

\section{Full MHD beyond the kinematic regime}  
In the following the results of the kinematic approach are compared with direct 
numerical simulations \cite{Bis93} as well as integrations of spectral transfer 
equations within closure models\cite{Pou78}, both at $Pr_m=1$ and with a kinetic 
energy about three orders of magnitude larger than the magnetic energy.  This 
provides insights into the feedback of the magnetic field on the flow via the 
Lorentz force density.  The question is whether the phenomena seen in the 
kinematic case, in particular the anticorrelation, will still be present. 
 
Both numerical experiments support the existence of a Kolmogorov-type  
scaling for the spectrum of the mean square magnetic potential,  
$A(k)\sim k^{-7/3}$ for  
$k<k_f$, in the inverse mean square flux cascade range. The Alfv\'{e}n 
effect causing an equipartion between kinetic and magnetic energy is 
subdominant. The spectral closure suggests $A(k)\sim k^{-7/2}$  
for the direct energy cascade range $k>k_f$.   
We start with a corresponding model spectrum 
for the  amplitudes $\langle|\psi_{\bf k}|^2\rangle$ (see Fig. \ref{p6}(a)) 
\end{multicols}
\begin{eqnarray}  \langle|\psi_{\bf k}|^2\rangle\!\sim\!  
\left\{   
\begin{array}{r@{\quad:\quad}l}   
k_f^{-7/6}k_1^{-13/3}k^3             & \frac{2\pi}{L}\le k\le k_1, \\   
k_f^{-7/6}k^{-4/3}                  & k_1<k\le k_f,\\   
k^{-5/2}                            & k_f<k\le k_d, \\   
k^{-5/2}\exp\left[-\left(\frac{k-k_d}{k_d}\right)^{2}\right]   
						& k_d<k\;,   
\end{array}  
\right.  
\label{modspec1}  
\end{eqnarray}  
\begin{multicols}{2}
where $k_d\sim\eta_d^{-1}$ is the wave number at the viscous cut off. 
The structure function is calculated as in section III: 
\begin{equation}   
\label{mod_struc1}   
D_2^{(\psi)}(r)=2\,\sum_k\,\langle|\psi_{\bf k}|^{2}\rangle\,   
(1-\mbox{J}_{0}(kr)) \,.   
\end{equation}   
The resulting structure function is shown in panel (b) of Fig.~\ref{p6}.  The 
inset gives $\zeta_2^{(\psi)}$ vs.  $r$ measured in units of $\eta_d$. 
Consequently $D_2^{(\psi)}(r)$ has to be taken now in units of 
$\epsilon_{\psi}\epsilon^{-1/2}\nu^{1/2}$.  The $r^2$--scaling for $r<\eta_d$ is 
followed by a range with $\zeta_2^{(\psi)}\approx \frac{4}{3}$ for $r>\eta_d$, 
which seems to be connected with the extended inverse cascade of the mean square 
magnetic potential.  Again structure function and correlation function are 
calculated by means of (\ref{laplace}) and (\ref{laplace1}), respectively. 
Consistency was checked by calculating $D_2^{(B)}(r)$ directly via $\langle|{\bf 
B}_{\bf k}|^2\rangle\sim k^2 \langle|\psi_{\bf k}|^2\rangle$ using a relation 
equivalent to (\ref{mod_struc1}).  The function $D_2^{(B)}(r)$ is shown in 
Fig.~\ref{p3} as the dash-dotted line.  We see that it reaches the constant 
saturation state for scales $r$ larger than in the kinematic case.  The 
differences between the kinematic and full dynamic situation are more pronounced 
in the correlation functions [see Fig.~\ref{p6}(c)].  The correlations decay 
slower with respect to $r$, following now a $r^{-2/3}$--law. 
 
Besides the different scalings we note that the anticorrelation  
has disappeared.  In the  
full dynamic case the magnetic field itself would prohibit the build-up of  
elongated current sheets (flux sheets) due to its feedback via the Lorentz 
force  density 
\begin{eqnarray}  
\label{Lorentz}  
{\bf f}_L=  
-{\bf \nabla}\left(\frac{B^2}{2\mu_0}\right)  
+\frac{1}{\mu_0}({\bf B}\cdot{\bf\nabla}){\bf B}\;.  
\end{eqnarray}  
The current sheets, if formed, become sensitive to several resistive 
instabilities such as the spontaneous growth of the tearing 
mode,\cite{FKR63,Mat86,Pol89} where the stored magnetic energy can be released into 
the plasma flow by means of magnetic reconnection.  In this process the Lorentz 
force density causes the plasma acceleration due to reconnected magnetic flux 
and prevents a further steepening of flux up to the dissipative scales.  Thus in 
addition to the condition $M>R_m$ the Reynolds number R has to be 
sufficiently low to avoid resistive MHD instabilities.  It was found that 
the onset of the tearing mode instability is determined by the   
Hartmann number $Ha=\sqrt{R_m R}=R \sqrt{Pr_m}$ and the width  
$\delta_{CS}$.\cite{Dah83,SeSch98}

\section{Summary}  
Our findings for the magnetic structure function as well as the corresponding  
correlation function in a two-dimensional 
MHD system can be summarized as  
follows.  
  
(1) For the kinematic regime of two-dimensional MHD turbulence ($M\gg 1$), the 
magnetic flux function $\psi$ can be treated as a passive scalar advected in a 
turbulent flow.  By means of the geometric scaling theory the second order 
structure function $D_2^{(\psi)}(r)$ was calculated over a wide range of scales 
$r$ and of magnetic Prandtl numbers $Pr_m$.  For this case a rather complete 
analysis is now possible by means of Eq. (\ref{fracdim}). 
 
(2) The second order magnetic structure function $D_2^{(B)}(r)$ and the 
corresponding magnetic correlation function $C_2^{(B)}(r)$ were calculated by 
exact analytical relations from $D_2^{(\psi)}(r)$.  The larger $Pr_m$ the 
smaller the scale $r$ where the structure function reaches the constant 
saturation.  The correlation function decays fast with $r^{-4/3}$ and shows an 
anticorrelation.  The latter is connected with the transition to the Batchelor 
regime and can be associated with an attempt of the system to concentrate its 
magnetic energy in strongly filamented structures.  This process is limited by 
the finite resistivity $\eta$.  The anticorrelation minimum was found to be 
independent of the magnetic Prandtl number. 
 
(3) The results for the kinematic approach were compared to results in the 
dynamic regime $M<R_m$.  The magnetic field correlations decay with $r^{-2/3}$, 
i.e.  slower than for the kinematic case.  This can be explained by the feedback 
of the magnetic field on the flow which prohibits the build-up of strongly 
filamented, elongated current sheets on small scales.  Therefore no  
anticorrelation was detected. 
 
While the discussion in this paper was limited to the 2-D case, a weak field 
regime without dynamo can also be identified in three dimensions.  In the 
experiments of Odier {\it et al.}\cite{Od98} a scaling of the magnetic field was observed that 
is very close to what one expects for a passive scalar.  Thus the stretching of 
the magnetic field does not seem to have a pronounced effect on the structure 
function.  Further investigations of this point seem worthwhile.

\acknowledgments  
Fruitful discussions with H. Fuchs and K.-H. R\"adler are 
gratefully acknowledged.

\appendix  
\section{Derivation of Equation (16)}  
  
The second order  
structure function $D_2^{(B)}(r)$ was decomposed in Eq. (\ref{definition_DC})  
into (the sum convention is used)  
\begin{eqnarray}  
D_2^{(B)}(r)&=&2\langle B_i^2 \rangle- 2 C_2^{(B)}(r),\nonumber\\  
		&=&2 R^{(B)}_{ii}(0)-2 R^{(B)}_{ii}({\bf r}),                                     
\end{eqnarray}   
where $R^{(B)}_{ij}({\bf r})=\langle B_i({\bf x})B_j({\bf x}+{\bf r})\rangle$ is  
the static magnetic field correlation tensor.  A detailed analysis  
of the properties of $R^{(B)}_{ij}({\bf r})$ can be found in  
Oughton {\it et al.}\cite{Oug97} Next  
we apply  
$\epsilon_{ijk}\epsilon_{ilm}=(\delta_{jl}\delta_{km}-\delta_{jm}\delta_{kl})$  
for ${\bf B}={\boldmath\nabla\times}{\bf A}$ in the autocorrelated part and get  
\begin{eqnarray}  
B_i^2&=&\epsilon_{ijk}\partial_{x_j} A_k   
	\epsilon_{ilm}\partial_{x_l} A_m,\nonumber\\  
     &=&(\partial_{x_j} A_m)^2-(\partial_{x_j} A_m \partial_{x_m} A_j).  
\end{eqnarray}   
The statistical average gives    
\begin{eqnarray}  
\langle B_i^2\rangle =\langle(\partial_{x_j} A_m)^2\rangle  
			   =\frac{\epsilon_A}{\eta},  
\end{eqnarray}   
where the second term vanishes due to homogeneity. The trace of the   
magnetic field correlation tensor  
can be decomposed in a similar way by using \cite{MonYag75}  
\begin{eqnarray}  
\langle B_i({\bf x}) B_i({\bf x}+{\bf r})\rangle&=&  
-\epsilon_{ijk}\epsilon_{ilm}\partial_{r_j}\partial_{r_l}R^{(A)}_{km}  
({\bf r}),\nonumber\\  
&=&-\partial^2_{r_l}R^{(A)}_{mm}({\bf r}),\nonumber\\  
&=&\frac{1}{2}\Delta D_2^{(A)}(r).  
\label{yaglom} 
\end{eqnarray}   
For the derivation the Coloumb gauge ${\boldmath\nabla}\cdot{\bf A}=0$   
has to be taken. Finally we get in 3-D  
\begin{eqnarray}   
D_2^{(B)}(r)=2\frac{\epsilon_A}{\eta}-\Delta D_2^{(A)}(r),  
\end{eqnarray}   
and it follows in 2-D with ${\bf A}=(0,0,\psi(x,y))$ Eq. (\ref{laplace})  
\begin{eqnarray}   
D_2^{(B)}(r)=2\frac{\epsilon_{\psi}}{\eta}-\Delta D_2^{(\psi)}(r).  
\end{eqnarray}   
 
The dimensionless form of (\ref{laplace}),   
$D_2^{(B)}=2Pr_m-\Delta D_2^{(\psi)}$, was  
solved numerically over a large range  
of scales $r$. We substitute the Laplacian in the following way  
\begin{eqnarray}   
\Delta D_2^{(\psi)}&=&  
\frac{1}{r}\partial_r\left[r\partial_r D_2^{(\psi)}\right],\nonumber\\  
&=&\frac{D_2^{(\psi)}}{r^2}\left[(\partial_{\ln r} g)^2  
					   +\partial_{\ln r}^2 g\right].  
\end{eqnarray}   
Here $g(\ln r)=\ln D_2^{(\psi)}(r)$ was set. The function $g$ is  
approximated  
with cubic splines in a least square fit using the Numerical  
Algorithms Group (NAG) library.

    
\pagebreak  
  
\begin{narrowtext}    
\begin{figure}  
\begin{center}  
\epsfig{file=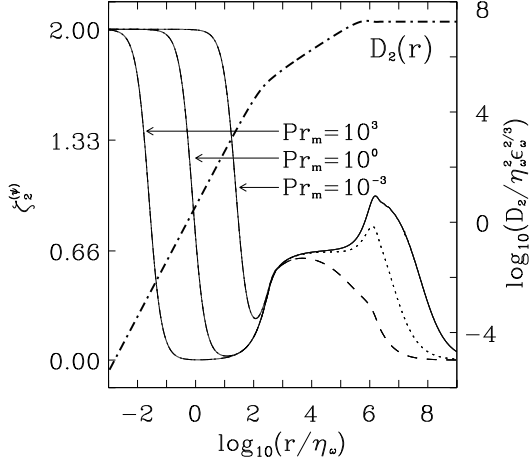,width=6.5cm,height=6.5cm}  
\end{center}  
\vspace{1cm}  
\caption{Velocity structure function 
$D_2(r)$ (dash-dotted line) and the scaling exponent 
$\zeta_2^{(\psi)}(r)$ of the structure function $D_2^{(\psi)}(r)$ for three 
values of the prefactor $\alpha=\epsilon_{\psi}\epsilon_{\omega}^{-1/3} 
\psi_{r.m.s.}^{-2}$ and three values of the magnetic Prandtl number.  The solid 
line stands for $\alpha=10^{-4}$, the dotted for $\alpha=10^{-3}$, and the 
dashed line for $\alpha=10^{-2}$.  For small distances $r$ the 
$\alpha$-dependence is subdominant and thus the curves for the same magnetic 
Prandtl number coincide.}   
\label{p1}  
\end{figure}  
\begin{figure}  
\begin{center}  
\epsfig{file=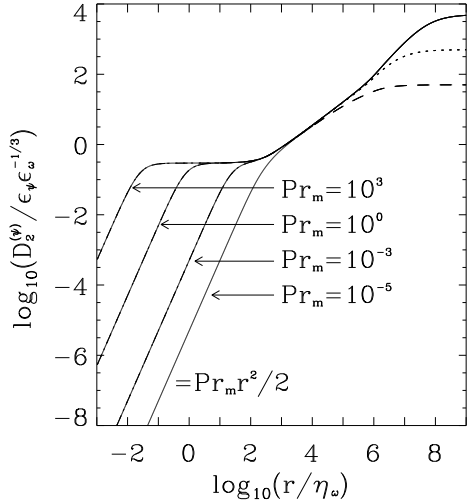,width=7.5cm,height=7.5cm}  
\end{center}  
\vspace{1cm}  
\caption{Structure function $D_2^{(\psi)}(r)$ with 
respect to $r$ for $\alpha$ as in Fig.~\ref{p1}.  Additionally the structure 
function for $Pr_m=10^{-5}$ for $\alpha=10^{-4}$
is shown which has no Batchelor plateau.  All 
linestyles are chosen as in Fig.  \ref{p1}.}  
\label{p2}  
\end{figure}
\vfill  
\begin{figure}  
\begin{center}  
\epsfig{file=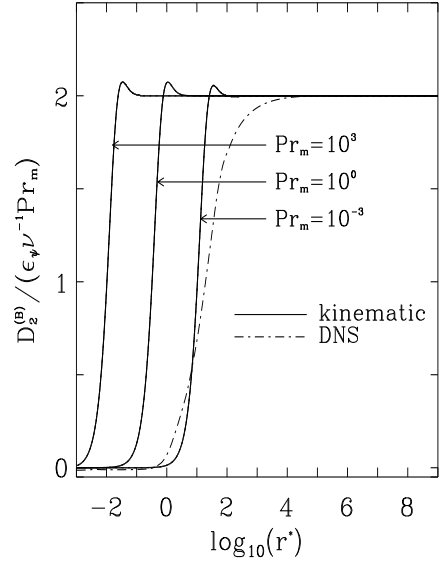,width=6cm,height=7cm}  
\end{center}  
\vspace{1cm}  
\caption{Normalized structure function\\ 
$D_2^{(B)}(r)/(\epsilon_{\psi}\nu^{-1}Pr_m)$ vs.  $r^{\ast}=r/\eta_{\omega}$ for 
$\alpha$ as in Fig.~\ref{p1}.  The curves for the different values of $\alpha$, 
but the same $Pr_m$, coincide.  Additionally, 
$D_2^{(B)}(r)/(\epsilon_{\psi}\nu^{-1})$ vs.  $r^{\ast}=r/\eta_d$ calculated 
from direct numerical simulations (DNS) at $Pr_m=1$  
is plotted as a dash-dotted line.}   
\label{p3}  
\end{figure}  
\begin{figure}  
\begin{center}  
\epsfig{file=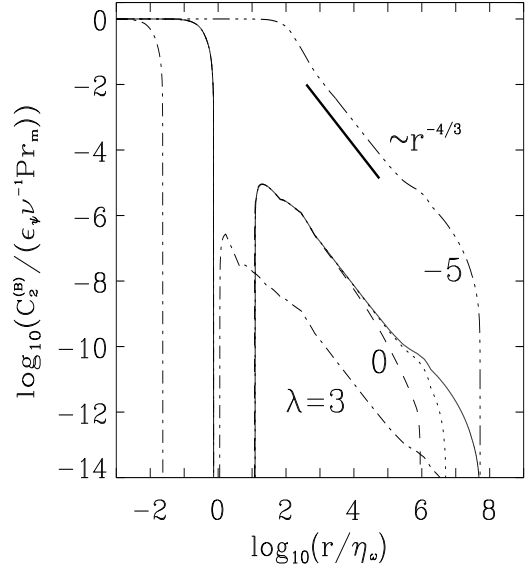,width=6cm,height=7cm}  
\end{center}  
\vspace{1cm}  
\caption{Normalized magnetic correlation function\\ 
$C_2^{(B)}(r)/(\epsilon_{\psi}\nu^{-1}Pr_m)$ vs.  $r/\eta_{\omega}$ for  
$Pr_m=10^{\lambda}$. 
For $Pr_m=1$ the correlation functions are shown for the three values of 
$\alpha$ as in Fig.~\ref{p1} using the same linestyles as therein.}   
\label{p4} 
\end{figure} 
\vfill 
\begin{figure}  
\begin{center}  
\epsfig{file=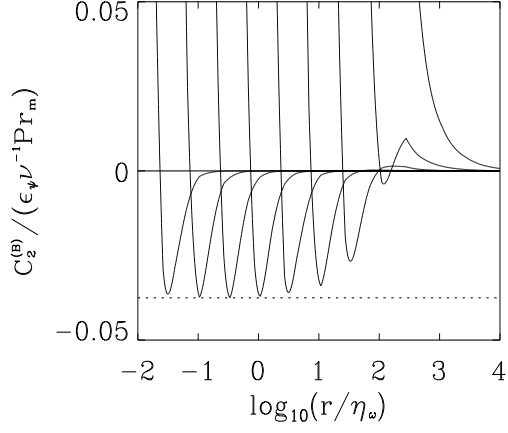,width=6cm,height=5cm}  
\end{center}  
\vspace{1cm}  
\caption{Anticorrelated part of 
$C_2^{(B)}(r)$ where the magnetic Prandtl number $Pr_m$  
decreases by a factor 10 from $10^{3}$ for the left most curve to $10^{-5}$ 
at the right end.  
The parameter $\alpha$ was set to $10^{-4}$.}   
\label{p5}  
\end{figure} 
\vfill 
\begin{figure}  
\begin{center}  
\epsfig{file=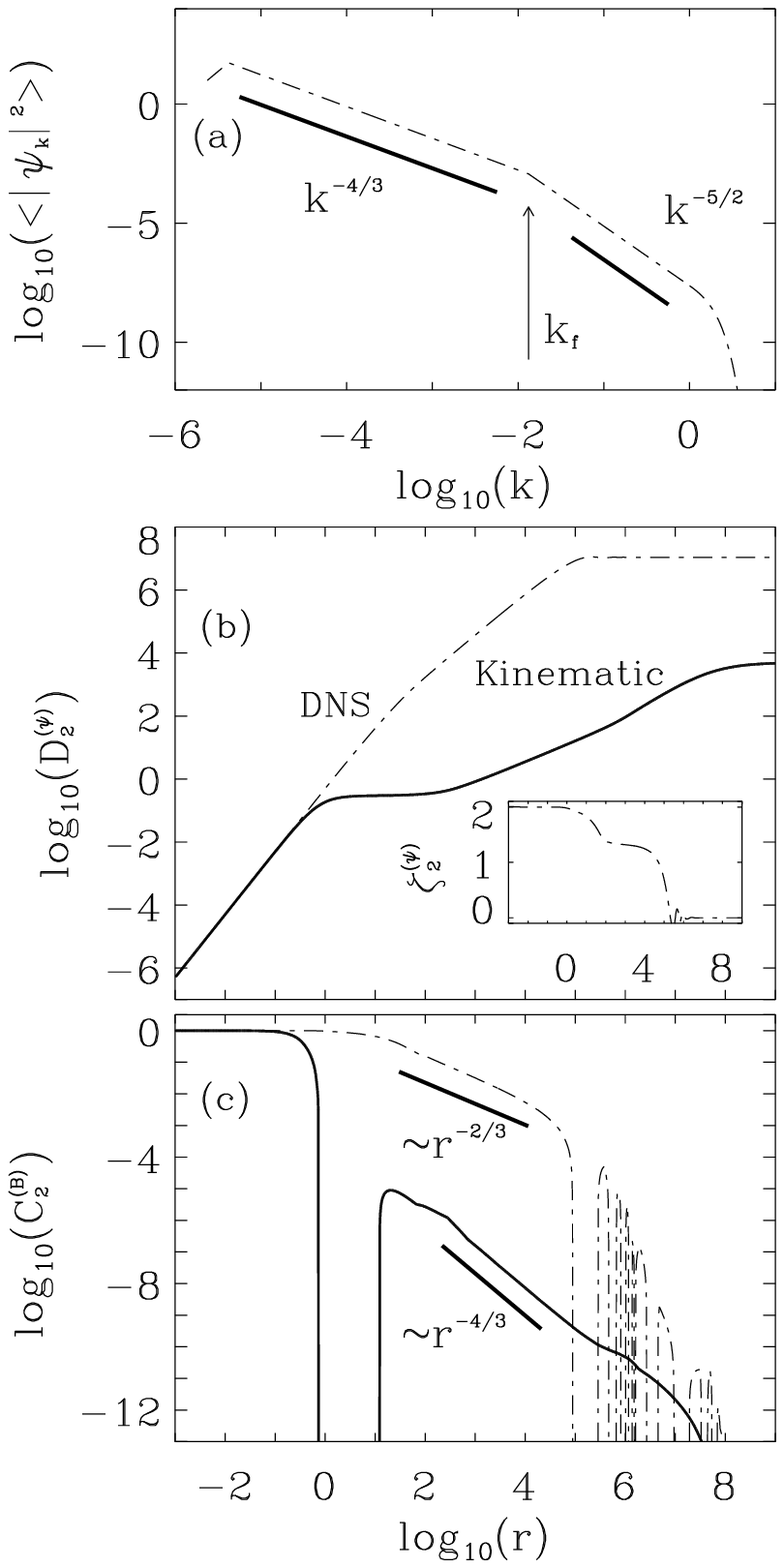,width=6.4cm,height=16cm}  
\end{center}
\vspace{1cm}  
\caption{(a):  Mean square potential spectrum 
$\langle|\psi_{\bf k}|^2\rangle$ vs.  $k$ resulting  
from full MHD simulation results. 
The extension of the two cascade ranges is identical to those of the velocity 
input spectrum for the kinematic calculations. 
(b):  Corresponding structure function 
$D_2^{(\psi)}(r)$ vs. $r$ plotted with the same linestyle as (a) and 
compared with its kinematic counterpart for $\alpha=10^{-4}$.  
The inset  shows the corresponding $\zeta_2^{(\psi)}(r)$.   
(c):  Both magnetic correlation 
functions are plotted corresponding to (b). All quantities are expressed  
in characteristical units and $Pr_m=1$ was taken.} 
\label{p6}  
\end{figure}  
\end{narrowtext}  
\end{multicols}

\begin{references}  
\bibitem{Bis94} D. Biskamp, {\it Nonlinear Magnetohydrodynamics}  
		    (Cambridge University Press, Cambridge, England, 1993).  
			 
\bibitem{Mar96} E. Marsch, C.-Y. Tu, and H. Rosenbauer,  
		    Ann. Geophysicae {\bf 14}, 259 (1996).  
		    
\bibitem{Kui95} J. Kuijpers, "Flares in Accretion Discs",  
		    in {\it Coronal Magnetic Energy Releases},  
		    A. O. Benz and A. Kr\"uger, eds. (Lecture Notes in Physics,  
		    Springer, Heidelberg, 1995), Vol. 444, pp. 135-158.  
    
\bibitem{Zel80} Ya. B. Zeldovich and A. A. Ruzmaikin, Sov. Phys. JETP {\bf 51},  
		    493 (1980).  
		     
\bibitem{Bis90} D. Biskamp and H. Welter, Phys. Fluids B {\bf 2}, 1787  
		    (1990).   
		     
\bibitem{Cat91} F. Cattaneo and S. I. Vainshtein, Astrophys. J. {\bf 376},  
		    L21 (1991).  
 
\bibitem{Con93} P. Constantin and I. Procaccia, Phys. Rev. E {\bf 47},   
		    3307 (1993);             
		    P. Constantin and I. Procaccia, Nonlinearity {\bf 7}, 1045  
		    (1994).  
 
\bibitem{Proc93} I. Procaccia and P.  Constantin, Europhys. Lett. {\bf 22},  
		689 (1993). 
		
\bibitem{EckSch98} B. Eckhardt and J. Schumacher, "Structure function of  
			 passive scalars in   
			 two-dimensional turbulence", submitted to Phys. Rev. E 
			 (1998).  
			 
\bibitem{Zel83} Ya. B. Zeldovich, {\it Magnetic Fields in Astrophysics} 
		    (Gordon and Breach, New York, U.S.A., 1983). 
 
\bibitem{Bis93a} D. Biskamp, Europhys. Lett. {\bf 21}, 563 (1993). 
  
\bibitem{Gra95}  R. Grauer and C. Marliani, Phys. Plasmas {\bf 2}, 41 (1995). 
 
\bibitem{GroMer92} S. Grossmann and P. Mertens, Z. Phys. B {\bf 88},  
			 105 (1992).  
 
\bibitem{Par97} J. Paret and P.Tabeling,  Phys. Rev. Lett. {\bf 79},  
		    4162 (1997); J. Paret and P.Tabeling, Phys. Fluids {\bf 10}, 
		    3126 (1998). 
			  
\bibitem{Rut98} M. A. Rutgers, Phys. Rev. Lett. {\bf 81}, 2244 (1998).               
 
\bibitem{FriSul84} U. Frisch and P. L. Sulem, Phys. Fluids {\bf 27},  
			 1921 (1984).   
				      
\bibitem{SmiYak93} L. M. Smith and V. Yakhot, Phys. Rev. Lett. {\bf 71},  
			 352 (1993).   
				      
\bibitem{Ben86}  R. Benzi, C. Paladin, S. Patarnello, P. Santangelo,  
		     and A. Vulpiani, J. Phys. A {\bf 19}, 3771 (1986).   
 
\bibitem{Tab99}  J. Paret, M.-C. Jullien, and P. Tabeling, 
		     ''Vorticity Statistics in the two-dimensional enstrophy  
		     cascade'', submitted to Phys. Rev. Lett. (1999).  
\bibitem{Bab87} A. Babiano, C. Basdevant, B. Legras, and R. Sadourny, 
		    J. Fluid Mech. {\bf 183}, 379 (1987). 
		     
\bibitem{Bas94} C. Basdevant and T. Philipovitch, Physica D {\bf 37}, 17 
		    (1994). 
 
\bibitem{Bat59} G. K. Batchelor, J. Fluid Mech. {\bf 5}, 113 (1959).   
  
\bibitem{Obu49} A. M. Obukhov, Izv. Akad. Nauk SSSR, Ser. Geog. Geofiz.    
		    {\bf 13}, 58 (1949);                 
		    S. Corrsin, J. Appl. Phys. {\bf 22}, 469 (1951).   
			 
\bibitem{Bat51} G. K. Batchelor, Proc. Camb. Phil. Soc. {\bf 47}, 359 (1951).   
		      
						  
\bibitem{Bis93}  D. Biskamp and U. Bremer, Phys. Rev. Lett {\bf 72}, 3819  
		     (1993).  
			 
\bibitem{Pou78}  A. Pouquet, U. Frisch, and J. L\'eorat, J. Fluid Mech.   
		     {\bf 77}, 321 (1976); A. Pouquet, J. Fluid Mech. {\bf 88},   
		     1 (1978).  
			    
\bibitem{FKR63} H.~P. Furth, J. Killeen, and M.~N. Rosenbluth,  
		    Phys. Fluids {\bf 6},  459 (1963). 
 
\bibitem{Mat86} W. H. Matthaeus and S. L. Lamkin, Phys. Fluids {\bf 29}, 2513  
		    (1986).  
 
\bibitem{Pol89} H. Politano, A. Pouquet, and P. L. Sulem,  
		    Phys. Fluids B {\bf 1}, 2330 (1989). 
		     
\bibitem{Dah83} R.~B. Dahlburg, T.~A. Zang, D. Montgomery, and  
		    M.~Y. Hussaini, Proc. Natl. Acad. Sci. USA {\bf 80},   
		    5798  (1983). 
 
\bibitem{SeSch98} N. Seehafer and J. Schumacher, Phys. Plasmas {\bf 5},   
			2363  (1998). 
 
\bibitem{Od98} P. Odier, J.-F. Pinton, and S. Fauve, Phys. Rev. E {\bf 58}, 
		   7397 (1998). 
		    
\bibitem{Oug97} S. Oughton, K.-H. R\"adler, and W. H. Matthaeus, Phys. Rev. E  
		    {\bf 56}, 2875 (1997).  
			 
\bibitem{MonYag75} A. S. Monin and A. M. Yaglom, {\it Statistical Fluid    
			 Mechanics}, (MIT Press, Cambridge, Massachusetts, 1975).   
\end{references}
\end{document}